\documentclass[pra,preprint,showpacs,preprintnumbers,amsmath,amssymb,floatfix]{revtex4}

\bibstyle{apsrev}
\usepackage{graphicx}
\usepackage{epstopdf}
\usepackage{bm}

\begin{document}
\title{Electron swaps and the stopping of protons by hydrogen}
\author{J. M. Rodr\'iguez Aguirre}
\email[]{juakcho@gmail.com}
\author{E. R. Custidiano}
\email[]{cernesto@exa.unne.edu.ar}
\affiliation{Departamento de F\'isica. Facultad de  Ciencias Exactas y Naturales, y Agrimensura. Universidad Nacional del Nordeste, Avda. Libertad 5600, 3400 Corrientes, Argentina}
\author{M. M. Jakas}
\email[]{mmateo@ull.es}
\affiliation{Departamento de F\'{\i}sica Fundamental y Experimental, Electr\'{o}nica y Sistemas. Universidad de La Laguna, 38205 La Laguna, Tenerife, Spain}

\date{\today}
\begin{abstract}
The relevance of the electronic swap in the stopping process of proton by hydrogen is investigated. To this end, the Classical Trajectory Monte-Carlo method is used to calculate the k-stopping cross-section, i.e. the stopping cross-section given the occurrence of k-swaps during the collision.  It is found that electron swaps can be used to label electron trajectories, as it seems to describe fairly well the extent of the electron-ion interaction during the collision.  Depending on the ion energy, the number of swaps entering the stopping cross section may vary. In the keV range the number of swaps can be as large as five or more, whereas, at larger energies, only three or less electron swaps may take place in collisions of relevance to stopping.
\end{abstract}
\maketitle

\section{Introduction}
\label{Introduction}

The so-called stopping power, or linear energy deposition for swift ions penetrating a material, i.e. $\delta$E/$\delta$x, is a relevant quantity in many areas of engineering and physics \cite{Johnson90}. It is normally defined as the mean energy-loss ($\delta$E) a projectile undergoes when passing through a $\delta$x-thick layer of material. From an atomistic point of view the stopping power is obtained using the equation:

\begin{equation}
\label{Eq:1}
\frac{\delta E}{\delta x} = -NS \, .	
\end{equation}

\noindent where $N$ is the atomic density of the material, and $S$ is the stopping cross-section, defined as 

\begin{equation}
\label{Eq:2}
S = \int dT \sigma (T) \, .	
\end{equation}

\noindent where $\sigma(T)$  is the differential scattering cross section as a function of the energy loss $T$.

The stopping cross section has been extensively studied both experimentally and theoretically \cite{Bohr13,Bethe30,Lindhard52,Firsov59,Johnson90,OlsonXX,Ziegler99}. From the theoretical side, following the seminal works of Bohr \cite{Bohr13}, Bethe \cite{Bethe30} and Lindhard \cite{Lindhard52}, a considerable progress has been achieved over the years. Furthermore, the increase of computer capability observed in the last decades brought on the possibility of numerically simulating the scattering processes, and triggered the interest of many researchers in re-studying the stopping at a level of detail not performed before. 

Computer simulations of the stopping are based on both quantum \cite{Schiwietz90,Cabrera02} and classical mechanics \cite{Tokesi99,Custidiano02}. On the classical side, the so-called Classical Trajectory Monte-Carlo method (CTMC) assumes that the nuclei and the electron are all classical particles. The simulation, therefore, first picks the initial state of the electron and the nuclei and, secondly solves the Newton's equation of motion for these three particles during the passage of the bombarding ion. The use of classical mechanics, however, is possibly one of the most serious shortcomings of this approach. Except for some cases where using classical mechanics is seemingly well justified, the results of these simulations may not be expected to be very accurate. Surprisingly enough, however, the results of the stopping cross sections calculated using the CTMC method showed that this approach is fairly accurate and that one may rely on it much more than expected \cite{Custidiano02,Olson03}. 

The concept of electronic swap was introduced in Ref.\cite{Homan94} in analyzing the charge exchange in ion collisions with Rydberg atoms. Although it will be explained in more detail in the following section, the electron swap refers to the passage of the electron from one nucleus to the other. Obviously, a swap may take place when the two nuclei are close enough so that electron is subjected to the forces of the two colliding nuclei. During the collision, the classical electron may realize none, one, two or more swaps. It is evident that the number of swaps tells one about the trajectory followed by the electron during the collision. Therefore, one may conceivably imagine that the number of swaps must be closely related to the stopping process as well. 

The present paper aims at analyzing the correlation between the electron swaps and stopping of protons by a classical hydrogen atom. To this end, a previously developed CTMC computer code \cite{Custidiano02,Custidiano05} is used, but modified so as to enable it to calculate of the stopping cross section for trajectories with the same number of swaps separately.  This paper is organized as follows: in Section \ref{Calculations} the computer simulations code is briefly described. Section \ref{Results} contains the results and discussions of the present calculations. Finally, summary and concluding remarks are offered in Section \ref{Summary}.

\section{Calculations.}
\label{Calculations}
\subsection{The CTMC method.}

The classical trajectory Monte-Carlo method (CTMC) assumes that both the electron and nuclei are classical, point-like, charged particles.  The evolution with time of the position and velocity of these particles is obtained by solving the Newton's equation of motion, where forces are calculated from the Coulomb potential acting between the electron and nuclei, and the initial position and velocity of the electron are randomly chosen using a certain statistical initial distribution function. 

An essential part of the CTMC calculation is the one trajectory event or, simply, a trajectory. It comprises the preparation of the initial state and the numerical calculation of the electron and nuclei trajectories up to a time when the state of the electron can be unambiguously determined. That is, the electron has to be exclusively excited, ionized or captured by the projectile.  

A trajectory is completely defined by the initial state of the electron and the impact parameter b of the incoming ion. Therefore, any quantity one would like to calculate should be obtained by taking average over a certain number of trajectories. A number that often depends on the statistical uncertainty one is prepared to accept. For example, the mean energy loss undergone by the incoming ion for a given impact parameter $b$ and bombarding energy $\varepsilon$, i.e. $\Delta E(b,\varepsilon)$, is obtained from calculating the average energy losses over $N$ trajectories events as:

\begin{equation}
\label{Eq:3}
\Delta E(b,\varepsilon) =\frac{\sum_{j=1}^{N} \Delta E_j(b,\varepsilon)}{N}
\end{equation}

where $\Delta E_j(b,E)$ is the energy transferred from the projectile to the electron and target nucleus in the $j$-th trajectory event, and which is often obtained from the difference between the initial and final energy of the electron. Similarly, one can obtain the stopping cross-section by calculating the integral:

\begin{equation}
\label{Eq:4}
S(\varepsilon) =2\pi\int_{0}^{\infty} db\: b\: \Delta E(b,E) 	\, .
\end{equation}

Using the CTMC, however, the integral above is calculated by taking average of $2\pi\: b\: \Delta E(b,E)$ over a number $M$ of impact parameter $b$, uniformly sampled between $0$ and $b_{max}$:   
 	
\begin{equation}
\label{Eq:5}
S(\varepsilon) \cong	2\pi b_{max} \frac{\sum_{j=1}^{M}b_j\: \Delta E_j(b_j,\varepsilon)}{M} \, ,
\end{equation}

\noindent where both $b_{max}$ and $M$ are chosen from a compromise between accuracy and computer time. Sometimes though, $b_{max}$ is obtained from physical arguments. This is the case of high bombarding energies where Eqs.(4-5) do not converge unless a finite $b_{max}$ is used. Normally $b_{max}$ is set to the so called adiabatic radius, defined as the impact parameter above which a real, quantum electron will no longer absorb energy from the passing charge \cite{Bohr13,Jackson62}. 

\subsection{Initial distribution.}

Unfortunately enough, the choice of the initial state of the electron is not as simple as one could possibly imagine. The initial position and velocity of the electron are sampled from a classical phase-space distribution function that should be similar to that of the target atom. However, this is not so simple, since one needs both the position and velocity, and quantum mechanics forbids the simultaneous knowledge of these two quantities. The introduction of some approximation is therefore unavoidable. 

For example, one can use the so-called microcanonical distribution \cite{Abrines66, Custidiano02}. In this approach, the initial state of the electron in the classical orbit is sampled from the distribution:

\begin{equation}
\label{Eq:6}
\rho (\mathbf{r},\mathbf{v}) = \frac{1}{8\pi^2}\:\delta \left(\frac{\bf{p}^2}{2m}-k_e\frac{e^2}{r}-E_0\right)\, ,
\end{equation}

\noindent where $\delta$ is the Dirac delta function \cite{Abramowitz}, $m$ is the mass of the electron, $e$ is the elementary charge, $k_e$ is the so-called Coulomb's constant, and  $E_0$ the energy of the ground state in the hydrogen atom, i.e. -12.6 eV. This distribution has the advantage of being stable over time and that every sampled electron will have the same energy. However, it is limited along the radial coordinate and this may cause some underestimation of the charge transfer and the stopping cross sections. 

\subsection{Swap number}

As was already mentioned, the electron is subjected to the combined Coulomb interaction of the two nuclei, i.e.: 

\begin{equation}
\label{Eq:7}
V_e(\mathbf{r}) = -k_e \frac{e^2}{\left|\mathbf{r}_e - \mathbf{r}_1\right|} - k_e \frac{e^2}{\left|\mathbf{r}_e - \mathbf{r}_2\right|}\, ,
\end{equation}

\noindent where $\mathbf{r}_e$ is the position of the electron and $\mathbf{r}_1$, $\mathbf{r}_2$ are those of the ion and target nuclei positions, respectively.

In order to identify under which nucleus the electron is much stronger interacting with, the space can be divided into two by a plane that is perpendicular to the straight line containing the two nuclei, that passes through the so-called \emph{saddle point} \cite{Gay96} defined by the expression:
 
\begin{equation}
\label{Eq:8}
\frac{1}{\left|\mathbf{r}_e - \mathbf{r}_1\right|} = \frac{1}{\left|\mathbf{r}_e - \mathbf{r}_2\right|}\, .
\end{equation}

\begin{figure}
\includegraphics[width=9.0cm]{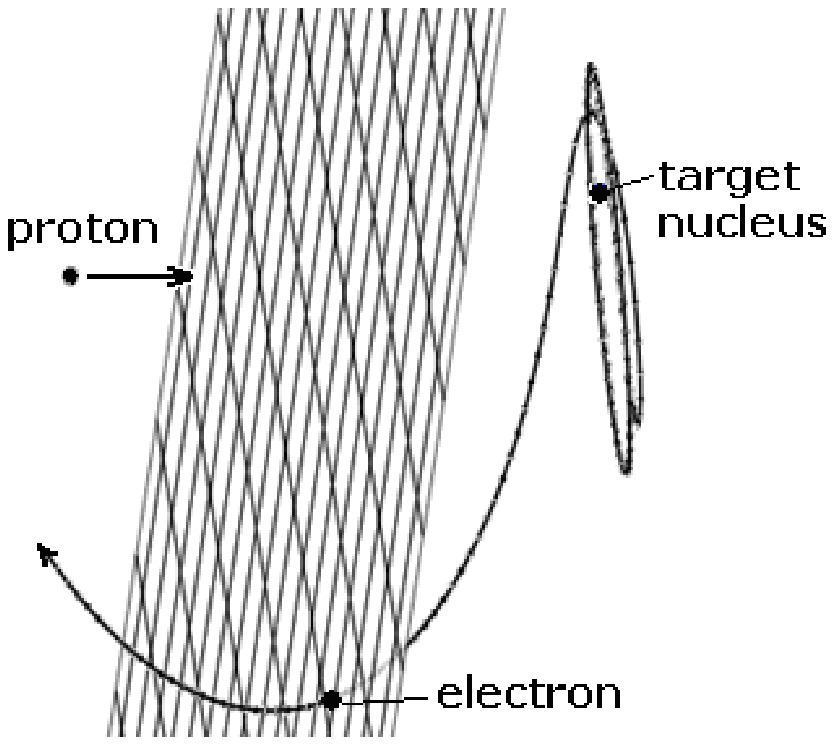}
\caption{\label{Fig:1} Path of the electron and position of the mid-plane (crossed stripe) for a $v$  = 1 atomic units (a.u.) proton (left) impinging on a Hydrogen atom (right).} 
\end{figure}

In the following, this plane will be referred to as the \emph{mid-plane} (see figure \ref{Fig:1}). One may thus introduce the swap number, i.e. $N_j$, as the number of crossings undergone by the electron through the mid-plane during the $j$-th trajectory event. Similarly, one can define the probability of having $k$-swaps after $N$ trajectories as:   

\begin{equation}
\label{Eq:9}
P_k(b,\varepsilon) =\frac{\sum_{j=1}^{N} \delta_{k,n_j}}{N}\, ,
\end{equation}

\noindent where $\delta_{i,j}$ is the Kroenecker delta function \cite{Abramowitz}. It can be easily verified that 
 
\begin{equation}
\label{Eq:10}
\sum_{k=0}^{\infty} P_k(b,\varepsilon) = 1\, ,
\end{equation}

\noindent And, similarly, the one can introduce the swap number cross-section as:

\begin{equation}
\label{Eq:11}	
\sigma(\varepsilon) = 2\pi \int_{0}^{\infty} db\:b\: P_k(b,\varepsilon)\, .
\end{equation}

\noindent This cross section is expected to be finite only for $k > 0$, since the integral above diverges for $k = 0$, because $P_0(b,\varepsilon) =1$ with $b \rightarrow \infty $. 

Similarly, one may define the $k$-swap stopping cross-section or, simply, the $k$-stopping, as 

\begin{equation}
\label{Eq:12}	
S_k(\varepsilon) = 2\pi \int_{0}^{\infty} db\: b\: \Delta E_k(b,\varepsilon)\, ,
\end{equation}

\noindent where $\Delta E_k(b,\varepsilon)$ represents the mean energy-loss associated with $k$-swap trajectories. 

It must be noticed that the total stopping cross section becomes 

\begin{equation}
\label{Eq:13}	
S(\varepsilon) = \sum_{k=0}^\infty S_k(\varepsilon) \, .
\end{equation}

It must be observed that, contrary to the swap-number cross section, the zero-swap stopping cross section can be readily calculated, since  $\Delta E_0(b,E)$ does not have any additional problem regarding convergence, apart from those pointed out above at the end of subsection 2a. 

In the following section the results of calculating the swap-number and $S_k$ for protons on Hydrogen will be presented and discussed.

\section{Results and discussions.}
\label{Results}

\begin{figure}
\includegraphics[width=12.0cm]{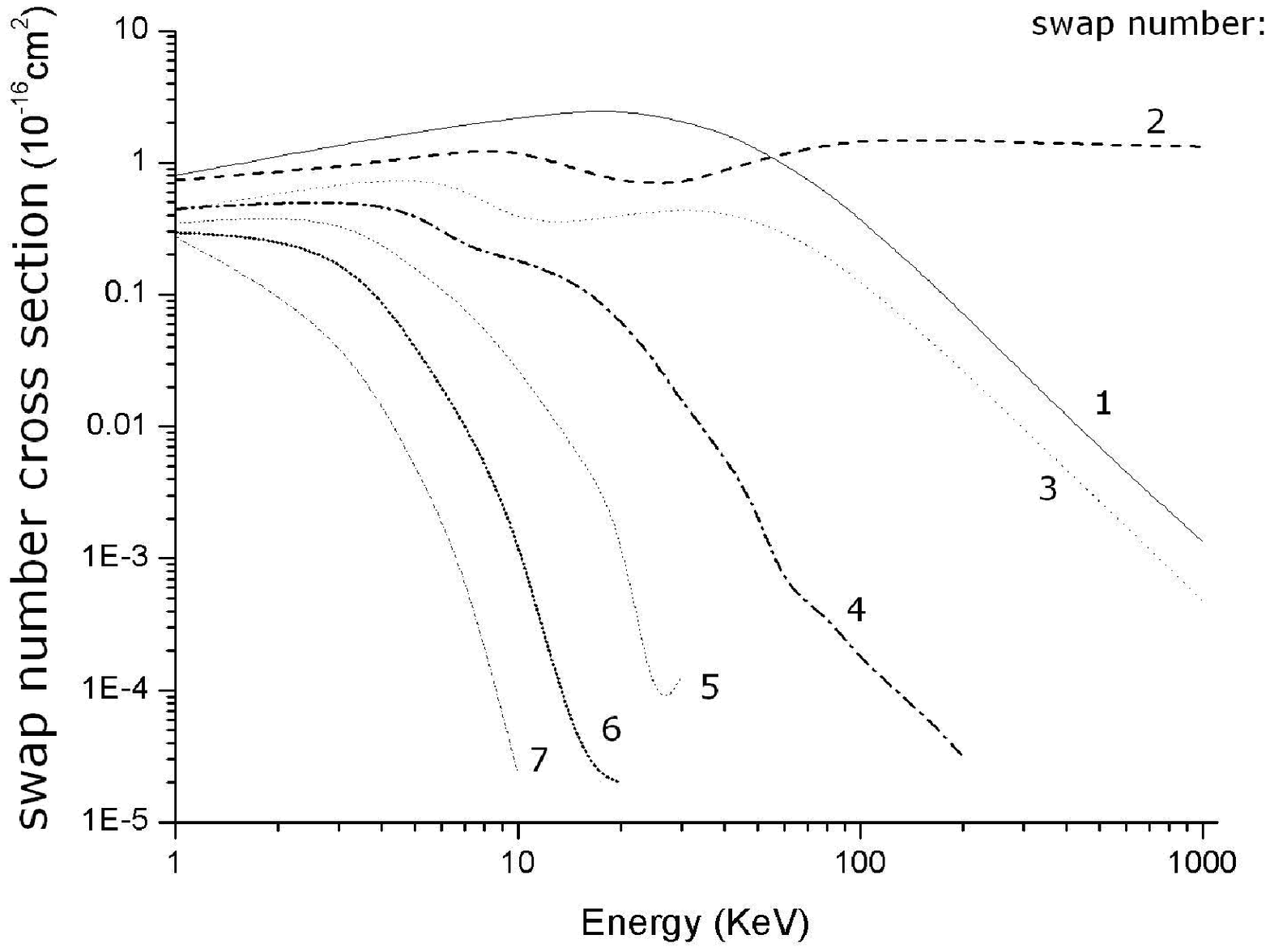}
\caption{\label{Fig:2} Swap-number cross-sections [see Eq.(\ref{Eq:11})] as functions of the bombarding energy.} 
\end{figure}

Figure \ref{Fig:2} shows the swap number cross-sections $\sigma_k(\varepsilon)$ as a function of the bombarding energy $\varepsilon$ and swap numbers ranging from 1 to 7. Notice that the (k = 0)-case does not appear since, as was previously commented, $\sigma_0(\varepsilon)$ cannot be calculated. In the first place one can see that, with the only exemption of $k$ = 2,  $\sigma_k$ is a decreasing function the swap number. Secondly, $\sigma_k(\varepsilon)$ appears to be a relatively smooth function of the energy for $k$ = 2, whereas it strongly depends on $\varepsilon$ for higher $k$-values. Third, it must be noticed that the $k$ = 2 case remains fairly constant over the entirely range of bombarding energy studied in this paper. This implies that the probability for the electron of performing one swing towards the projectile and coming back to the target nucleus is nearly the same irrespective of the bombarding energy.  It is evident that with a decrease of the bombarding energy the number of swaps increases. This agrees with the commonly accepted picture that, at low energy, the electron performs several turns around the two nuclei during the collision. 

The fact that $\sigma_2$ remains at nearly a constant value with an increase of the ion energy is not at all unexpected. As a matter of fact, excepting for an exceedingly small number of head-on encounters between the projectile and the electron, the latter is practically at rest compared to the projectile velocity and so, the plane used to count the swaps may pass either zero or two times over the electron, irrespective of the ion energy. The case of zero-swap however is not shown in the figure due to its ill-definition, as was previously mentioned.

\begin{figure}
\includegraphics[width=12.0cm]{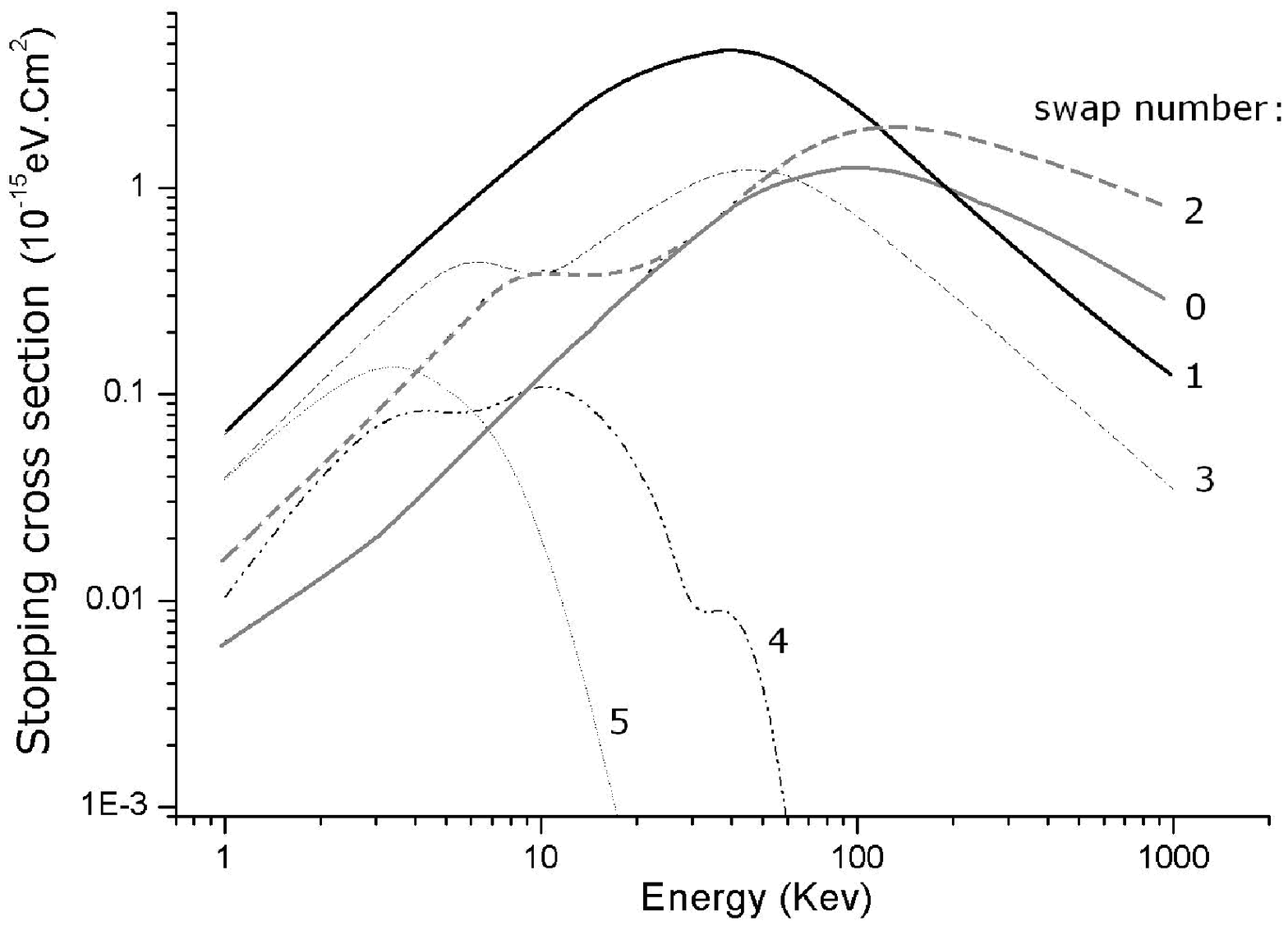}
\caption{\label{Fig:3} $k$-swap stopping cross-sections [see Eq.(\ref{Eq:12})] as functions of the bombarding energy.} 
\end{figure}

The results of calculating the k-stopping cross-section [see Eq.(\ref{Eq:12})] for $k$ = 0 to 5 are depicted in figure \ref{Fig:3}. There one can see that for bombarding energy lower than 100 keV, $S_1$ dominates, whereas, for larger energies, $S_2$ becomes the largest of them all. It must be noticed that both $S_4$ and $S_5$ drop down very rapidly at energies greater than 10 keV, becoming negligible small for energies greater than 100 keV. This is due to the fact that swap numbers greater than 3 are fairly unlikely events for energies larger than 10 keV, as was already observed in figure \ref{Fig:2}.  

\begin{figure}
\includegraphics[width=12.0cm]{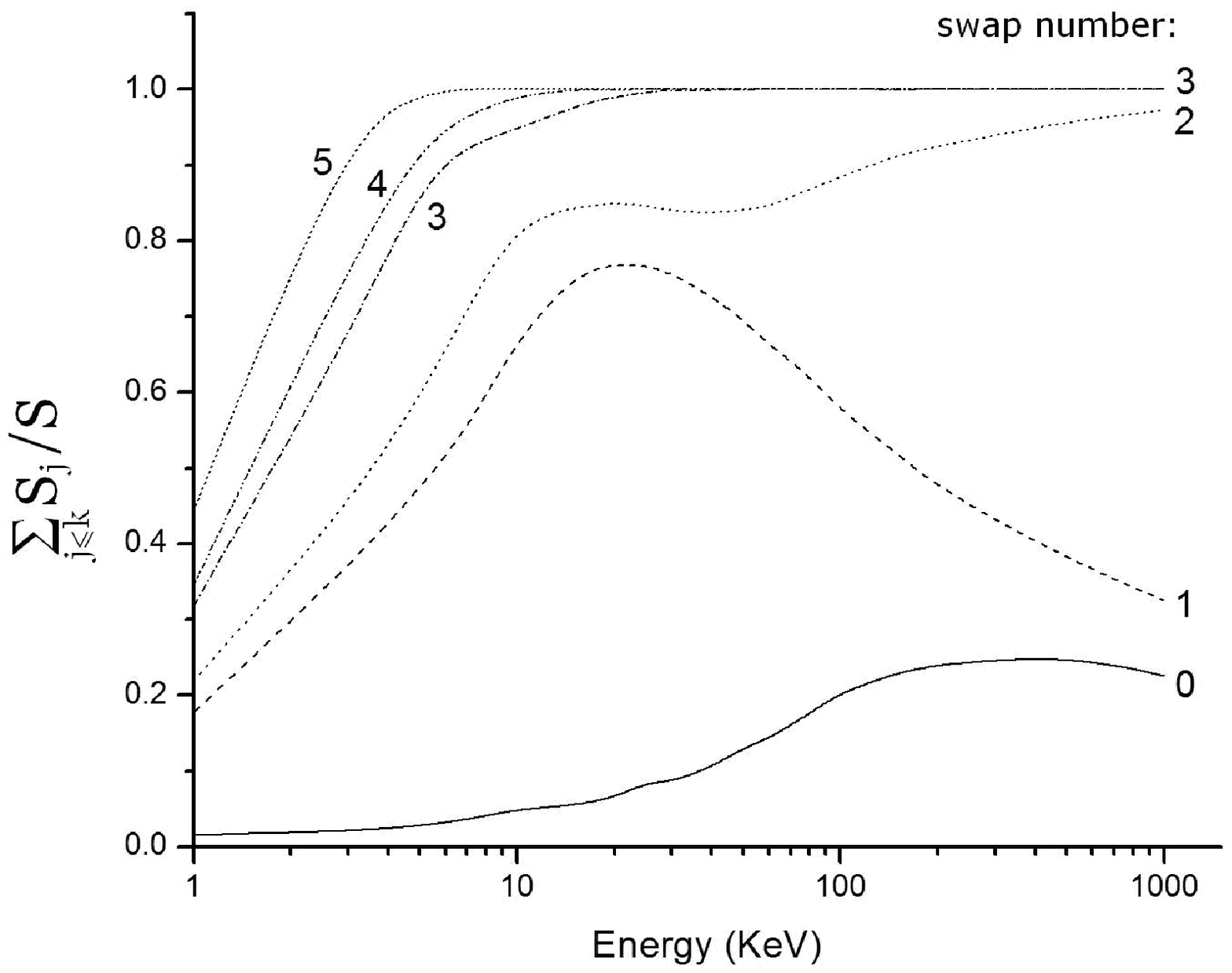}
\caption{\label{Fig:4} Accumulated $k$-swap stopping as functions of the bombarding energy. $S_j$ is obtained from Eq.(\ref{Eq:12}) and $S$ is the total stopping cross section, i.e. $S =\sum_{j=0}^{\infty} S_j$.} 
\end{figure}

Figure \ref{Fig:4} shows the sum of the first k-swap stopping, normalized to the total stopping cross section. This plot shows in a much clear manner the contribution of the first $k$-stopping to the total one. As one can readily see, for energies smaller than 3 keV the number of swaps associated with the stopping process is fairly larger than 5. As the bombarding energy increases this number decreases, too. However, at large energies the contribution of one- and zero-swap collisions decrease and that of the two-swap collisions becomes the dominant contribution. 

\begin{figure}
\includegraphics[width=12.0cm]{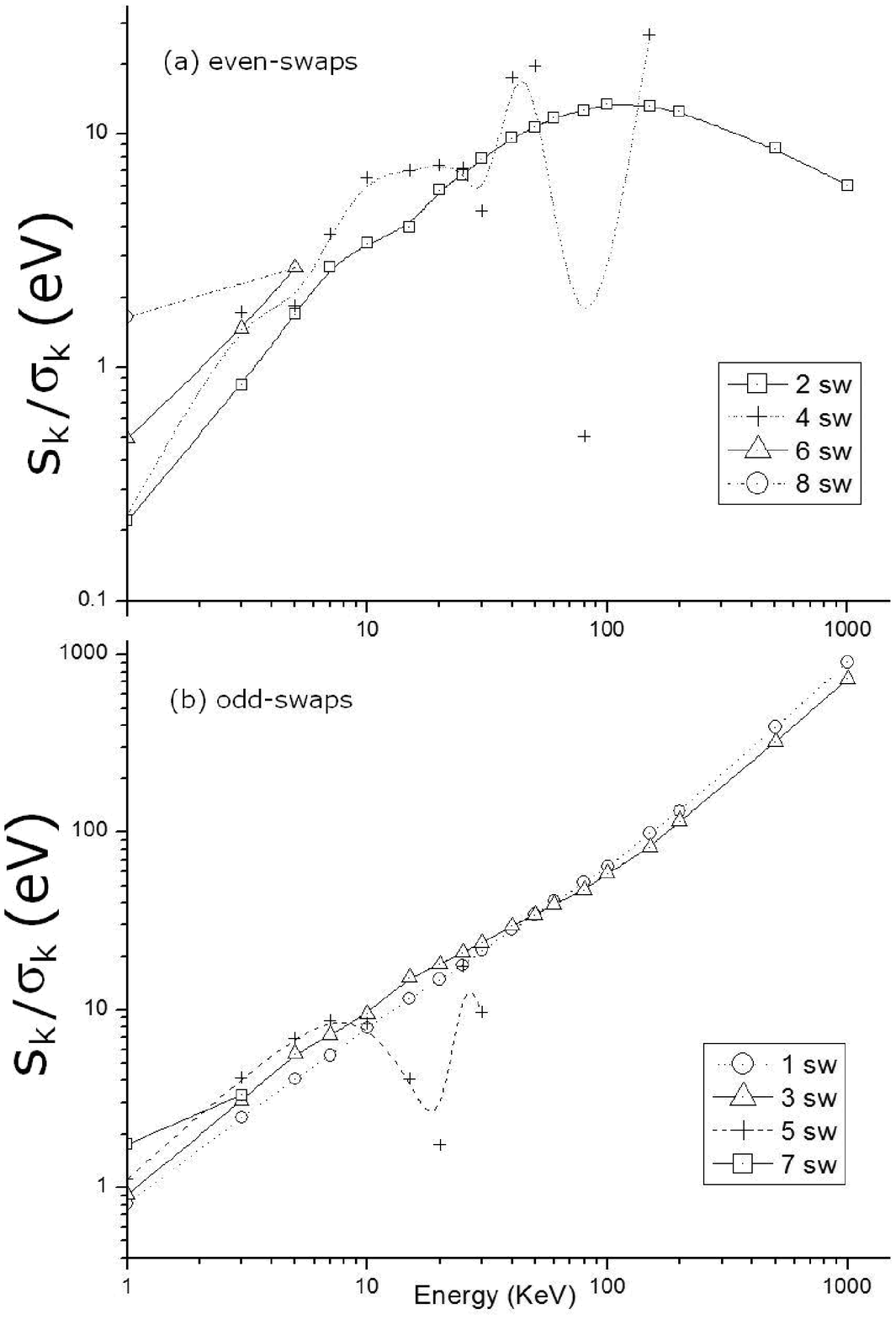}
\caption{\label{Fig:5} Mean energy-loss per collision, defined as the ratio between the $k$-stopping and the corresponding $k$-swap cross section.  Results are grouped by (a) even- and (b) odd-swap collisions, respectively.} 
\end{figure}

One may obtain the mean energy-loss in a $k$-swap collision by dividing the $k$-stopping by the $k$-swap cross section. The results are shown in the figure \ref{Fig:5}, where even and odd swaps collision are plotted separately. As one can readily see, there is a clear difference between even- and odd-swaps collisions. In the first place, the mean energy-loss corresponding to odd-swaps are greater than those of even-swaps. Secondly, the mean energy-losses of odd-swaps collisions are monotonously increasing functions of the projectile energy, whereas those of the even swaps are bell-shaped with maximum around 100 keV.
 
\begin{figure}
\includegraphics[width=12.0cm]{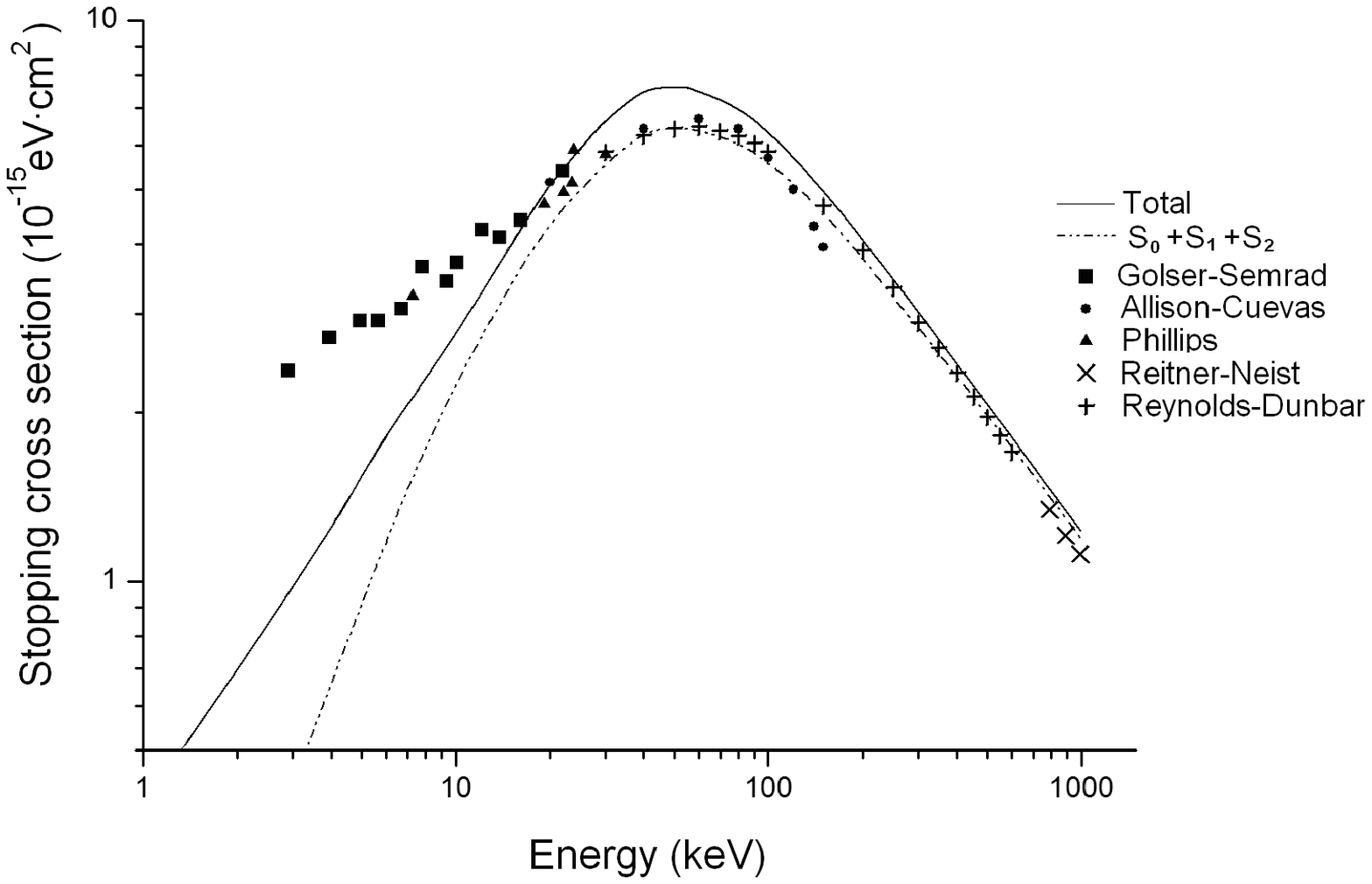}
\caption{\label{Fig:6} Comparison between present calculations (lines) and experimental data (symbols). Refs.\cite{Phillips53, Reynolds53, Allison62, Reitner90, Golser92}.} 
\end{figure}

In figure \ref{Fig:6}, the total stopping cross-section calculated in this paper is compared with experimental data from Refs. \cite{Phillips53,Reynolds53,Allison62,Reitner90,Golser92}. As was previously discussed \cite{Custidiano02,Custidiano05}, the present computer simulations reproduce experiments fairly well over a wide range of bombarding energy. Discrepancies are observed below 10 keV, however, such differences may be conceivably attributed to the fact that molecular, not atomic hydrogen targets are used in those experiments. With regard to the electron swaps, it must be noticed that by adding the first three swap-stopping cross sections, i.e. 0, 1 and 2, one may account, reasonably well, for the total stopping.     

\section{Summary and concluding remarks.}
\label{Summary}

By using the Classical Trajectory Monte-Carlo method, the correlation between the so-called electron swaps and the stopping cross section is investigated.  The electron swap, a concept introduced in Ref.\cite{Homan94} in connection with the charge exchange during collision of ions with Rydberg atoms, relates to the passage of the classical electron through the ideal surface that separates the attraction field of the ion and that of the target nucleus (see text for a more precise definition). According to the results in this paper, the swap-number, or number of swaps performed by the electron during the collision appears to be a useful parameter in the stopping calculations.  In the first place, it turns out that the swap-number appropriately labels the electron trajectory, in the sense that it does convey information about the extent the electron has interacted with the incoming ion during the collision.  At bombarding energies lower than 1 keV the stopping seems to be associated with a fairly large number of swaps, whereas, with an increase of the bombarding energy the number of swaps relevant to stopping decreases. For bombarding energies larger than 100 keV, however, two-swap events appear to be strongly correlated with the stopping process, while both the zero and one-swap events become less relevant. Curiously enough, the mean energy-loss associated with odd-swap events seems to be a linear function of the bombarding energy, whereas those of the even-swaps are not only smaller but also bell-shaped with a maximum around 100 keV.  

\acknowledgments
One of the Authors (JMRA) acknowledges support from the Consejo Nacional de Investigaciones Científicas y Técnicas (CONICET), Argentina.


\end{document}